\documentclass[11pt]{article}
\usepackage{amsmath,amssymb}
\usepackage[dvipdfmx]{graphicx}

\newcommand{\ba}{\begin{alignat}{3}}

\newcommand{\pa}{\partial}

\begin{document}

\begin{titlepage}
\begin{flushright}
\end{flushright}
\begin{center}
  \vspace{3cm}
  {\bf \Large Black Hole and Fuzzy Objects \\[0.2cm] in BFSS Matrix Model}
  \\  \vspace{2cm}
  Yoshifumi Hyakutake
   \\ \vspace{1cm}
   {\it College of Science, Ibaraki University \\
   Bunkyo 2-1-1, Mito, Ibaraki 310-8512, Japan}
\end{center}

\vspace{2cm}
\begin{abstract}
We analyze fuzzy configurations of D0-branes in BFSS matrix model as microstates of black hole.
Fuzzy configurations of D0-branes consist of localized fuzzy objects in 3 spatial directions 
which are smeared into 6 internal directions.
Since the solutions are time dependent, these are non-BPS configurations and have internal energy 
compared with static case.
Especially we examine the smeared fuzzy sphere in the BFSS matrix model,
which will correspond to the microstate of the charged black hole in 4 dimensions,
and compare the effective potential in that background with the result obtained by the near horizon 
geometry of the black 0-brane.
The qualitative features of two descriptions agree with each other, thus
we expect the smeared fuzzy sphere corresponds to one of the microstates of the charged black hole.
We also examine the smeared fuzzy cylinder which will correspond to the flat space time.
\end{abstract}

\end{titlepage}

\setlength{\baselineskip}{0.65cm}


\section{Introduction}


Superstring theory is the promising candidate for the theory of quantum gravity.
In the superstring theory, D-branes play important roles to investigate
both gauge theory and gravity theory\cite{Polchinski:1995mt}.
Some of black holes are realized as bound states of D-branes, and
microstates of the black hole are constructed from the gauge theory on the D-branes\cite{Strominger:1996sh}.
This shows that the quantum nature of the gravity can be captured by analysing corresponding quantum field theory.

Actually quantum aspects of black hole are investigated considerably via matrix models,
which are non-perturbative formulation of superstring theories\cite{Banks:1996vh}-\cite{Itoyama:1998et}. 
In 1996, Banks, Fischler, Shenker and Susskind proposed a non-perturbative formulation of 
M-theory (BFSS matrix model)\cite{Banks:1996vh}.
This theory is obtained by dimensional reduction of 10 dimensional super Yang-Mills theory to quantum mechanics, 
which is identified with the effective action for multiple D0-branes\cite{deWit:1988ig}.
It is remarkable that although BFSS matrix model is the quantum mechanical system, it can reproduce the
gravitational force between two D0-branes\cite{Banks:1996vh,Douglas:1996yp,Becker:1997wh,Becker:1997xw}\footnote{
Originally BFSS proposed to take the size of matrices infinite. 
In ref.~\cite{Susskind:1997cw}, finite case was proposed as discretized light cone quantization of M-theory.
See ref.~\cite{Taylor:2001vb}, for example, as a review of the BFSS matrix model.}.
Hence BFSS matrix model captures the nature of the gravity, and
it is possible to investigate black hole physics in detail.

In fact, there are several works which construct Schwarzschild black hole in various dimensions from 
the BFSS matrix model\cite{Banks:1997hz}-\cite{Horowitz:1997fr}.
Thermodynamics of the black hole is reproduced qualitatively from the BFSS matrix model 
in ref.~\cite{Banks:1997hz}-\cite{Halyo:1997wj}, and instability of black string is examined in ref.~\cite{Horowitz:1997fr}.
Especially a fuzzy sphere configuration of D0-branes is used to describe the black hole in ref.~\cite{Kabat:1997im}.
In that paper, the effective potential for a test D0-brane in the background of the fuzzy sphere is evaluated at one-loop level, 
and it agrees with the result of the gravity side qualitatively.

In 1997, Maldacena conjectured the gauge/gravity correspondence\cite{Maldacena:1997re}, and it is confirmed 
that physical quantities in the gauge theory, such as correlation functions, are consistently calculated 
from the gravity side\cite{Gubser:1998bc,Witten:1998qj}.
From the viewpoint of this conjecture, it is natural to regard that the BFSS matrix model corresponds to 
the near horizon geometry of black 0-brane\cite{Itzhaki:1998dd}.
Since the gauge/gravity correspondence is a kind of weak/strong coupling duality, it is hard to test the conjecture analytically.
However, recently there are several numerical tests of this conjecture 
for the thermal system of D0-branes\cite{Anagnostopoulos:2007fw}-\cite{Berkowitz:2016jlq}.
Especially numerical study for the black hole geometry is considered in ref.~\cite{Rinaldi:2017mjl}.

If the gauge/gravity correspondence is correct even for the non-supersymmetric system, 
it is important to construct the black hole geometry in the BFSS matrix model.
Actually thermodynamic features of near extremal black branes are reproduced qualitatively 
by analysing corresponding super Yang-Mills theory\cite{Smilga:2008bt}-\cite{Morita:2014ypa}.
In ref.~\cite{Hyakutake:2016sig}, a smeared black 0-brane solution and its thermodynamic properties are investigated.
The black 0-brane is smeared along 6 internal directions, so it becomes a charged black hole in 4 dimensions.
Then it is interesting to consider corresponding configurations in the BFSS matrix model.
In this paper, we propose that fuzzy configurations of D0-branes,
which are time dependent fuzzy objects smeared along 6 internal directions, 
correspond to the microstates of the smeared black 0-brane in the near horizon limit.
These fuzzy objects are bound states of D0-branes and oscillating around the origin of 3 spatial directions.

The organization of this paper is as follows.
In section 2, we review the BFSS matrix model and construct time dependent fuzzy objects,
including fuzzy sphere and fuzzy cylinder.
In section 3, we review the one-loop effective potential for the test D0-brane in the background of fuzzy configuration.
In section 4, the effective potentials between fuzzy objects, such as fuzzy sphere and cylinder, 
and the test D0-brane are calculated explicitly.
We compare properties of these effective potentials with the result obtained by the gravity side in section 5.
Section 6 is devoted to conclusion and discussion.


\section{Time Dependent Fuzzy Objects in BFSS Matrix Model}


In this section, we consider time dependent configurations of $N$ D0-branes in BFSS matrix model, 
which will correspond to the microstates of the black hole.
Since we are interested in the black hole in 4 dimensional spacetime, we make
a fuzzy object in 3 spatial directions via D0-branes and smear it along remaining 6 internal directions.
The equations of motion for the fuzzy object are expressed by simultaneous nonlinear differential equations
and solutions have nontrivial time dependence in general.
We show numerical plot for $N=2$ case which represents a nontrivial bound state of two D0-branes,
and then explain analytic solutions for oscillating fuzzy sphere and fuzzy cylinder.

Let us consider the BFSS matrix model which describes the dynamics of multiple D0-branes\cite{Banks:1996vh}.
The action for D0-branes can be obtained by the dimensional reduction of 10 dimensional $\mathcal{N}=1$ super Yang-Mills theory
to 1 dimensional super quantum mechanics\cite{deWit:1988ig}. 
The supermultiplet consists of a gauge field $A_t$, 9 scalar fields $\Phi_i$ and a Majorana-Weyl fermion $\theta$.
These are expressed by $N \times N$ matrices, and the action for multiple D0-branes is given by
\begin{alignat}{3}
  \mathcal{S}_0 &= \frac{1}{g_\text{YM}^2} \int dt \, \text{tr} \Big(
  \frac{1}{2} D_t \Phi_i D_t \Phi^i + \frac{1}{4} [\Phi_i,\Phi_j]^2
  + \frac{i}{2} \theta^T D_t \theta 
  + \frac{1}{2} \theta^T \gamma^i [\Phi_i,\theta] \Big), \label{eq:actD0}
\end{alignat}
where $i,j=1,\cdots,9$, $D_t = \partial_t - i [A_t,\;]$ and $\gamma^i$ are $16 \times 16$ matrices.
The coupling constant $g_\text{YM}$ has mass dimension $3/2$.
Note that there are 9 scalar fields $X_i = 2\pi \ell_s^2 \Phi_i$, and diagonal element of $X_i$ 
corresponds to a position of each D0-brane in the $x_i$ direction.
Thus the size of the matrices $N$ is equal to the number of D0-branes.
By setting $\theta = 0$, the equations of motion for $\Phi^i$ and $A_t$ become
\begin{alignat}{3}
  D_t( D_t{\Phi}_i) = [\Phi^j,[\Phi_i,\Phi_j]], \qquad [\Phi^i, D_t \Phi_i] = 0. \label{eq:EOMmat}
\end{alignat}
The equation of motion for $\theta$ is trivially satisfied when $\theta = 0$.

Let us examine the eq.~(\ref{eq:EOMmat}) to construct the fuzzy object
which would correspond to the microstate of the black hole in 4 dimensions.
First of all, we set $A_t = 0$ and choose $\Phi_a (a=1,2,3)$ as
\begin{alignat}{3}
  (\Phi_1^\text{bg})_{mn} &= \tfrac{1}{2} \rho_m(t) \delta_{m+1,n} + \tfrac{1}{2} \rho_n(t) \delta_{m,n+1}, \notag
  \\
  (\Phi_2^\text{bg})_{mn} &= - \tfrac{i}{2}  \rho_m(t) \delta_{m+1,n} + \tfrac{i}{2}  \rho_n(t) \delta_{m,n+1}, \label{eq:fuzzy}
  \\
  (\Phi_3^\text{bg})_{mn} &= z_m(t) \delta_{m,n}. \notag
\end{alignat}
Here $\rho_m(t)$ and $z_m(t)$ are functions of temporal coordinate $t$, and $m,n = 1, \cdots, N$.
The superscript bg stands for the background.
This ansatz represents the fuzzy object in 3 dimensions which is axially symmetric
around $x^3$ direction\cite{Kim:1998et,Hyakutake:2001kn}.
Roughly speaking, $\rho_m$ represents the extension of the fuzzy object from the origin on the $x_3=z_m$ plane.
Therefore the fuzzy object makes an axially symmetric surface in 3 dimensions, and it carries a dielectric D2-brane charge.
Remaining 6 scalars $\Phi_u (u=4,\cdots,9)$ are chosen to be diagonal so that the fuzzy object
is smeared along 6 spatial directions. 

Let us substitute the ansatz (\ref{eq:fuzzy}) into the equations of motion (\ref{eq:EOMmat}).
The differential equations for $\rho_m$ and $z_m$ are written as
\begin{alignat}{3}
  \ddot{\rho}_m &= \big\{ \tfrac{1}{2} (\rho_{m+1}^2 - 2 \rho_m^2 + \rho_{m-1}^2) - (z_{m+1} - z_m)^2 \big\} \rho_m, & \qquad &(m=1,\cdots,N\!-\!1) \notag
  \\
  \ddot{z}_m &= \rho_m^2 (z_{m+1} - z_m) - \rho_{m-1}^2 (z_m - z_{m-1}), & \qquad &(m=1,\cdots,N) \label{eq:EOMcz}
\end{alignat}
where $\rho_0 = \rho_N = z_0 = z_{N+1} = 0$.
Note that the second equation in the eq.~(\ref{eq:EOMmat}) is automatically satisfied.
Since the differential equations (\ref{eq:EOMcz}) are nonlinear, in general it is impossible to obtain analytic solutions.
From the energy conservation, however, we see that
\begin{alignat}{3}
  E &= \frac{1}{g_\text{YM}^2} \text{tr} \Big( \frac{1}{2} \dot{\Phi}_i \dot{\Phi}^i - \frac{1}{4} [\Phi_i,\Phi_j]^2 \Big) \notag
  \\
  &= \frac{1}{g_\text{YM}^2} \sum_{m=1}^N \Big\{ \tfrac{1}{2} (\dot{\rho}_m)^2 + \tfrac{1}{2} (\dot{z}_m)^2 
  + \tfrac{1}{8} (\rho_m^2 - \rho_{m-1}^2)^2 + \tfrac{1}{2} (z_{m+1} - z_m)^2 \rho_m^2 \Big\}.
\end{alignat}
This shows that the ranges of $\rho_m$ are finite, and those of $(z_m - z_{m-1})$ are also finite if $\rho_m \neq 0$ for all $m$.
If $\rho_m=0$ for some $m$, the representation is reducible and it represents two or more fuzzy objects.
Since separations of those fuzzy objects should be finite, again $(z_m - z_{m-1})$ are also finite.
From the second equation of (\ref{eq:EOMcz}), we note that $\sum_{m=1}^N \ddot{z}_m = 0$, 
so we set the center of mass $\sum_{m=1}^N z_m = 0$ without loss of generality.
Thus the D0-branes are bounded around the origin in 3 spatial directions.

\begin{figure}[htb]
\begin{center}
\includegraphics[keepaspectratio,width=10cm,height=6cm]{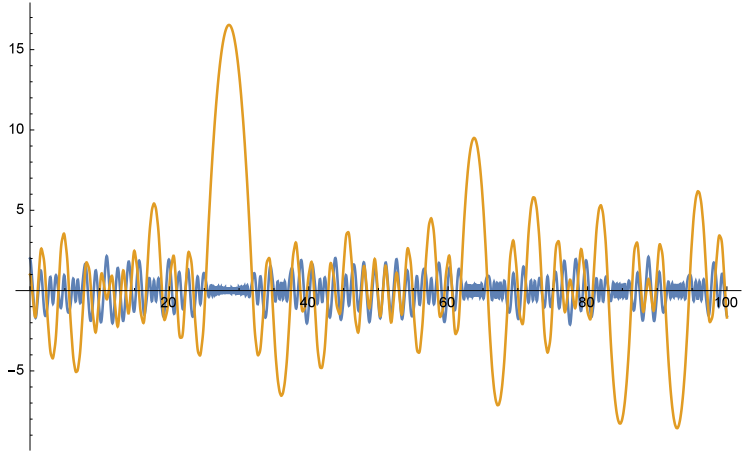}
\caption{Plots of $\rho_1(t)$ (blue) and $z_1(t)$ (yellow). Initial conditions are chosen as $\rho_1(0) = z_1(0) = 1$
and $\dot{\rho}_1(0) = \dot{z}_1(0) = 0$.}
\label{fig:czN=2}
\end{center}
\end{figure}

Now let us examine three types of solutions of the eq.~(\ref{eq:EOMcz}).
The first example is the case of $N=2$. Here we set $z_2 = - z_1$. 
Then differential equations for $\rho_1$ and $z_1$ are written as
\begin{alignat}{3}
  &\ddot{\rho}_1 = - ( \rho_1^2 + 4 z_1^2 ) \rho_1, \qquad \ddot{z}_1 = - 2 \rho_1^2 z_1, \label{eq:EOMN=2}
\end{alignat}
and the energy is given by
\begin{alignat}{3}
  E_{N=2} &= \frac{1}{g_\text{YM}^2} \big\{ \tfrac{1}{2}  (\dot{\rho}_1)^2 + (\dot{z}_1)^2 + \tfrac{1}{4} \rho_1^4 
  + 2 z_1^2 \rho_1^2 \big\} . \label{eq:EOMN=2ene}
\end{alignat}
Since $\rho_1=0$ corresponds to freely moving two D0-branes, we ignore this case.
Then $\rho_1$ and $z_1$ interact in a nontrivial way, but two D0-branes make a bound state around the origin in 3 dimensions.
Plots of $\rho_1(t)$ and $z_1(t)$ are shown in fig.~\ref{fig:czN=2}.
This shows that the dynamics of the fuzzy object is complicated even for $N=2$.

The second example is a fuzzy sphere which oscillates around the origin in 3 dimensions. 
The explicit forms of 3 scalar fields $\Phi_a^\text{bg}$ are given by the eq.~(\ref{eq:fuzzy}) with
\begin{alignat}{3}
  \rho_m = \tilde{r}(t) \sqrt{m(N-m)}, \qquad z_m = \frac{\tilde{r}(t)}{2} (N - 2m + 1). \label{eq:fs}
\end{alignat}
We also set $A_t^\text{bg}=0$ and 6 scalar fields $\Phi_u^\text{bg}$ to be diagonal.
$|\tilde{r}(t)|$ corresponds to the size of the fuzzy sphere, and the sign of $\tilde{r}(t)$ is related to the 
orientation of the sphere.
Tilde is used to clarify that the quantity has mass dimension.
Note that $\Phi_a^\text{bg}$ can be expressed as\cite{Madore:1991bw}
\begin{alignat}{3}
  \Phi_a^\text{bg} = \tilde{r}(t) \frac{\Sigma_a}{2}, \qquad 
  \Big[ \frac{\Sigma_a}{2},\frac{\Sigma_b}{2} \Big] = i\epsilon_{abc} \frac{\Sigma^c}{2},
\end{alignat}
where $a,b,c=1,2,3$. 
$\Sigma_a/2$ are $N$ dimensional irreducible representation of SU(2) Lie algebra,
which satisfy $\Sigma_1^2 + \Sigma_2^2 + \Sigma_3^2 = (N^2-1) {\bf 1}_N$.
Inserting the eq.~(\ref{eq:fs}) into the equations of motion (\ref{eq:EOMcz}), 
we obtain a simple equation for $\tilde{r}(t)$ as follows.
\begin{alignat}{3}
  \ddot{\tilde{r}} = - 2 \tilde{r}^3. \label{eq:f}
\end{alignat}
The above equation corresponds to the classical motion for a particle 
which is periodically moving in the quartic potential\cite{Collins:1976eg}.
And the solution is described by using Jacobi's elliptic function sn as
\begin{alignat}{3}
  \tilde{r}(t) = c_1 \,\text{sn}(c_1 t + c_2,-1),
\end{alignat}
where $c_1 (>0)$ and $c_2$ are integral constants.
$\tilde{r}(t)$ oscillates between $-c_1$ and $c_1$.
The radius of the fuzzy sphere is estimated as
\begin{alignat}{3}
  R_\text{sph} = \sqrt{\frac{1}{N} \text{tr} \big(X_1^2 + X_2^2 + X_3^2 \big)}
  = \pi \ell_s^2 \tilde{r}(t) \sqrt{N^2 - 1},
\end{alignat}
and the maximum value of the radius is given by $\pi \ell_s^2 c_1 \sqrt{N^2-1}$. 
In other words, $c_1$ is linearly related to the size of the fuzzy sphere.
Another constant $c_2$ can be fixed so that the radius of the fuzzy sphere becomes maximum at $t=0$.
The energy of the fuzzy sphere is estimated as
\begin{alignat}{3}
  E_\text{sph} &= \frac{N(N^2-1)}{8 g_\text{YM}^2} \big( \dot{\tilde{r}}^2 + \tilde{r}^4 \big) 
  = \frac{N(N^2-1)}{8 g_\text{YM}^2} c_1^4. \label{eq:Esph}
\end{alignat}
Note that this is the internal energy of $N$ D0-branes.
If we trust the classical solution naively, the fuzzy sphere oscillates around the origin in 3 dimensions.
And it carries a dielectric D2-brane or anti D2-brane charge, depending on the orientation of the sphere.
In actual, the fuzzy sphere interacts with the closed strings and it will lose the internal energy during the oscillation.

The third example is an oscillating fuzzy cylinder.
The fuzzy cylinder is homogeneously extending along $x_3$ axis and its circular cross section is oscillating on the $(x_1,x_2)$-plane.
Since the length of the fuzzy cylinder is infinite, the size of the matrices $N$ should be infinite.
The explicit forms of 3 scalar fields $\Phi_a^\text{bg}$ are given by the eq.~(\ref{eq:fuzzy}) with
\begin{alignat}{3}
  \rho_m = \tilde{\rho}(t), \qquad z_m = - \tilde{l} m, \label{eq:fc}
\end{alignat}
where $m$ takes integer value and $\tilde{l}$ is a typical mass scale of the fuzzy cylinder.
In the matrix representation, $\Phi_a^\text{bg}$ are expressed as
\begin{alignat}{3}
  &\Phi_1^\text{bg} = \tilde{\rho}(t) \, \Xi_1, \qquad
  \Phi_2^\text{bg} = \tilde{\rho}(t) \, \Xi_2, \qquad
  \Phi_3^\text{bg} = \tilde{l} \, \Xi_3, \notag
  \\
  &[\Xi_1, \Xi_2] = 0, \qquad\;\;
  [\Xi_2, \Xi_3] = i \Xi_1, \qquad
  [\Xi_3, \Xi_1] = i \Xi_2.
\end{alignat}
We set $A_t^\text{bg}=0$ and 6 scalar fields $\Phi_u^\text{bg}$ to be diagonal as the case of the fuzzy sphere.
Inserting the eq.~(\ref{eq:fc}) into the equations of motion (\ref{eq:EOMcz}), 
we obtain a simple equation for $\tilde{\rho}(t)$ as follows.
\begin{alignat}{3}
  \ddot{\tilde{\rho}} = - \tilde{l}^2 \tilde{\rho}. \label{eq:rho}
\end{alignat}
And the solution becomes
\begin{alignat}{3}
  \tilde{\rho}(t) &= c_3 \cos (\tilde{l} t + c_4).
\end{alignat}
Thus the fuzzy cylinder shrinks and expands like a harmonic oscillator.
The radius of the fuzzy cylinder is estimated as
\begin{alignat}{3}
  R_\text{cyl} &= \sqrt{\frac{1}{N} \text{tr} \big(X_1^2 + X_2^2 \big)} = 2 \pi \ell_s^2\tilde{\rho}(t),
\end{alignat}
and the internal energy is given by
\begin{alignat}{3}
  E_\text{cyl} &= \frac{N}{2 g_\text{YM}^2} \big( \dot{\tilde{\rho}}^2 + \tilde{l}^2 \tilde{\rho}^2 \big) 
  = \frac{N}{2 g_\text{YM}^2} \tilde{l}^2 c_3^2. \label{eq:Ecyl}
\end{alignat}
The fuzzy cylinder oscillates around the $x^3$ axis with carrying a dielectric D2-brane or anti D2-brane charge.
In actual, the fuzzy cylinder interacts with the closed strings and it will lose the internal energy during the oscillation.


\section{Fluctuations around Smeared Fuzzy Background}


In the previous section, we have constructed the fuzzy objects, such as fuzzy sphere and cylinder, in 3 dimensions.
In order to identify the fuzzy object with the microstate of black hole in 4 dimension, 
we need to smear it into spatial $x_u (u=4,\cdots,9)$ directions.
Thus we compactify $x_u$ with radius $R_u$, and put a copy of fuzzy object on each position of 
$(x_4,\cdots,x_9) = (2\pi R_4 n_4/Z_4, \cdots, 2\pi R_9 n_9/Z_9)$.
Here $Z_u$ are some integers and $n_u$ runs from $1$ to $Z_u$,
and there are $Z=\prod_{u=4}^9 Z_u$ copies of the fuzzy object.
By using the eq.~(\ref{eq:fuzzy}), 9 scalar fields for the smeared fuzzy object are represented as
\begin{alignat}{3}
  \Phi_a &= \Phi_a^\text{bg} \otimes {\bf 1}_Z, &\qquad &a=1,2,3, \notag
  \\
  \Phi_u &= {\bf 1}_N \otimes P_u, &\qquad &u=4,\cdots,9. \label{eq:fuzzybg}
\end{alignat}
Here $P_u$ are $Z \times Z$ diagonal matrices of the forms,
\begin{alignat}{3}
  P_u &= {\bf 1}_{Z_4} \otimes \cdots \otimes \frac{2 \pi \tilde{R}_u}{Z_u}
  \begin{pmatrix} 
    1 & & \\
    & \ddots & \\
    & & Z_u
  \end{pmatrix}
  \otimes \cdots \otimes {\bf 1}_{Z_9},
\end{alignat}
where $\tilde{R}_u = R_u/(2\pi \ell_s^2)$.
Thus each diagonal component composes a vector $p_u = 2\pi \tilde{R}_u n_u/Z_u$.
The fuzzy object is completely smeared when we take $Z_u \to \infty$.
The energy of the fuzzy objects are slightly modified due to the presence of $P_u$.
For examples, the internal energies of the fuzzy sphere and fuzzy cylinder are estimated as
\begin{alignat}{3}
  E_\text{sph} &= \frac{(NZ)^2(N^2-1)}{8 \lambda} c_1^4, \qquad
  E_\text{cyl} = \frac{(NZ)^2}{2 \lambda} \tilde{l}^2 c_3^2. \label{eq:Esphcyl}
\end{alignat}
where $\lambda = g_\text{YM}^2 NZ$ is the 't Hooft coupling constant.
The number of copies $Z$ goes to infinity, and $N$ is also infinite for the fuzzy cylinder.

Below we closely follow the ref.~\cite{Kabat:1997im} to evaluate the effective potential 
between the smeared fuzzy object and a test D0-brane.
In order to execute this, we start from the Euclidean action of the BFSS matrix model 
with the background field method.
We decompose the scalar fields as $\Phi_i = B_i + Y_i$.
Here $B_i$ are background fields and $Y_i$ are fluctuations.
As was solved in the previous section, backgrounds of the gauge field and the Majorana-Weyl fermion 
are set to be zero.
By adding gauge fixing and ghost terms, the action is given by
\begin{alignat}{3}
  \mathcal{S}_\text{E} &= \frac{1}{2 g_\text{YM}^2} \int d\tau \, \text{tr} \Big(
  D_\tau \Phi_i D_\tau \Phi^i - \frac{1}{2} [\Phi_i,\Phi_j]^2 
  + \theta^T D_\tau \theta - \theta^T \gamma^i [\Phi_i,\theta] \notag
  \\
  &\qquad\qquad\qquad\quad\;\;
  + (\dot{A}_\tau - i [B_i,Y^i])^2 
  - i \dot{\bar{C}} D_\tau C - [B_i, \bar{C}] D^i C \Big). \label{eq:SEu}
\end{alignat}
Here $\tau = i t$ is the Euclidean time and dot is the derivative with respect to $\tau$.
The explicit expressions for the background fields $B_i = (B_a, B_u)$ are written as
\begin{alignat}{3}
  B_a = \begin{pmatrix} \Phi_a^\text{bg} \otimes {\bf 1}_Z & 0 \\ 0 & \tilde{x}_a \end{pmatrix}, \qquad
  B_u = \begin{pmatrix} {\bf 1}_N \otimes P_u & 0 \\ 0 & 0 \end{pmatrix}. \label{eq:bg}
\end{alignat}
The first $NZ \times NZ$ block diagonal represents the smeared fuzzy object.
The second $1 \times 1$ component does the test D0-brane,
and $x_a = (2\pi\ell_s^2) \tilde{x}_a$ represents its position in 3 directions.
The tilde for $\tilde{x}_a$ is used to clarify that the quantity has mass dimension.
Below we assume that the test D0-brane is moving very slowly and it is reasonable to 
neglect the time dependence of $\tilde{x}_a$.

Let us consider fluctuations around the background (\ref{eq:bg}).
Since we are interested in the effective potential between the fuzzy object and the test D0-brane,
we only introduce the fluctuations of off-diagonal parts.
\begin{alignat}{3}
  &A_\tau = \begin{pmatrix} 0 & a(\tau) \\ a(\tau)^\dagger & 0 \end{pmatrix}, \quad&
  &\Phi_i = B_i + \begin{pmatrix} 0 & \phi_i(\tau) \\ \phi_i(\tau)^\dagger & 0 \end{pmatrix}, \notag
  \\
  &\theta = \begin{pmatrix} 0 & \psi(\tau) \\ \psi(\tau)^\dagger & 0 \end{pmatrix}, \quad&
  &C = \begin{pmatrix} 0 & c(\tau) \\ c^\dagger(\tau) & 0 \end{pmatrix}, \quad
  \bar{C} = \begin{pmatrix} 0 & \bar{c}(\tau) \\ \bar{c}^\dagger(\tau) & 0 \end{pmatrix}.
\end{alignat}
We substitute the above ansatz into the Euclidean action (\ref{eq:SEu}), and expand it up to the quadratic order 
of the fluctuations. Then the mass squared terms for 10 bosons $(a,\phi_i)$, the Majorana-Weyl fermion $\theta$ and 2 ghosts $c, \bar{c}$ 
are obtained as follows\cite{Kabat:1997im}.
\begin{alignat}{3}
  \Omega_\text{b}^2 &= K^2 \, {\bf 1}_{10} + M_{\text{b}}, \qquad
  \Omega_\text{f}^2 &= K^2 \, {\bf 1}_{16} + M_{\text{f}}, \qquad
  \Omega_\text{g}^2 &= K^2 \, {\bf 1}_{2}.
\end{alignat}
Here the diagonal parts of the mass squared terms have the same structure $K^2 = K_i K^i$,
which is $(NZ) \times (NZ)$ matrix. 
The explicit expressions for $K_i = (K_a, K_u)$ and $K^2$ are given by
\begin{alignat}{3}
  K_a &= Q_a \otimes {\bf 1}_Z , \qquad
  K_u = {\bf 1}_N \otimes P_u, \qquad
  K^2 = Q^2 \otimes {\bf 1}_Z + {\bf 1}_N \otimes P^2,
\end{alignat}
where 
\begin{alignat}{3}
  Q_a &= \Phi_a^\text{bg} - \tilde{x}_a \, {\bf 1}_{N}, \qquad
  Q^2 = (\Phi^\text{bg})^2 + \tilde{x}^2 {\bf 1}_N - 2 \tilde{x}^a \Phi_a^\text{bg}, \label{eq:Q_a}
\end{alignat}
and $\tilde{x}^2 = \tilde{x}_a \tilde{x}^a$, $(\Phi^\text{bg})^2 = \Phi_a^\text{bg} \Phi^\text{bg}{}^a$ and $P^2 = P_u P^u$.
On the other hand, off-diagonal parts of the mass squared terms, $M_{\text{b}}$ and $M_{\text{f}}$, are written as
\begin{alignat}{3}
  M_{\text{b}} &= 2i \begin{pmatrix} 0 & \dot{K}_j \\ - \dot{K}_i & -i[K_i,K_j] \end{pmatrix}
  \equiv 2i \begin{pmatrix} 0 & F_{\tau j} \\ F_{i\tau} & F_{ij} \end{pmatrix}, 
  \\
  M_{\text{f}} &= \gamma^i \dot{K}_i + \frac{1}{2} \gamma^{ij} [K_i,K_j]
  \equiv \frac{i}{2} \gamma^{\mu\nu} F_{\mu\nu}. \notag
\end{alignat}
In the second line, we introduced `10 dimensional' gamma matrices $\gamma^\mu \,(\mu=\tau,1,\cdots,9)$ 
and defined $\gamma^{\tau i} \equiv -i \gamma^i$.
Note that each $F_{\mu\nu}$ is $(NZ) \times (NZ)$ matrix.

Now we are ready to evaluate the effective potential at 1-loop level. 
The formula for the effective potential is given by
\begin{alignat}{3}
  V_\text{eff} &= \text{tr}_\text{b} (\Omega_\text{b}) - \frac{1}{2} \text{tr}_\text{f} (\Omega_\text{f}) 
  - \text{tr}_\text{g} (\Omega_\text{g}) \label{eq:Veff}
  \\
  &= - \frac{1}{2\sqrt{\pi}} \int_0^\infty \frac{d\ell}{\ell^{3/2}} \text{tr}_\text{b} \big( e^{-\ell \Omega_\text{b}^2} \big)
  + \frac{1}{4\sqrt{\pi}} \int_0^\infty \frac{d\ell}{\ell^{3/2}} \text{tr}_\text{f} (e^{-\ell \Omega_\text{f}^2} \big)
  + \frac{1}{2\sqrt{\pi}} \int_0^\infty \frac{d\ell}{\ell^{3/2}} \text{tr}_\text{g} (e^{-\ell \Omega_\text{g}^2} \big). \notag
\end{alignat}
Each term in the above can be evaluated perturbatively in the interaction picture.
In order to evaluate $e^{-\ell \Omega^2} = e^{-\ell (K^2 + M)}$, let us define 
$U(\ell) \equiv e^{\ell K^2} e^{-\ell \Omega^2}$ and $M(\ell) \equiv e^{\ell K^2} M e^{-\ell K^2}$.
The $U(\ell)$ satisfies a differential equation
$\frac{dU(\ell)}{d\ell} = - M(\ell) U(\ell)$,
and it can be solved as
\begin{alignat}{3}
  U(\ell) = {\bf 1} - \int_0^\ell d\ell_1 M(\ell_1) + \int_0^\ell d\ell_1 M(\ell_1) \int_0^{\ell_1} d\ell_2 M(\ell_2) - \cdots.
\end{alignat}
Thus $e^{-\ell \Omega^2} = e^{-\ell K^2} U(\ell)$ is expanded as
\begin{alignat}{3}
  e^{-\ell \Omega^2} 
  = e^{-\ell K^2} - \int_0^\ell d\ell_1 \, e^{-\ell K^2} M(\ell_1) 
  + \int_0^\ell d\ell_1 \int_0^{\ell_1} d\ell_2 \, e^{-\ell K^2} M(\ell_1) M(\ell_2) - \cdots. \label{eq:int}
\end{alignat}
Now it is possible to evaluate the effective action (\ref{eq:Veff}) order by order by employing the eq.~(\ref{eq:int}).
After some calculations, we see that terms up to the order of $M^3$ vanish because of the underlying supersymmetry.
The non-trivial contribution arises from the order of $M^4$, and the result is given by\cite{Chepelev:1997fk,Kabat:1997im}
\begin{alignat}{3}
  V_{\text{eff}} \big|_{M^4} &= 
  - \frac{1}{2\sqrt{\pi}} \int_0^\infty \frac{d\ell}{\ell^{3/2}} \int_0^\ell d\ell_1 \int_0^{\ell_1} d\ell_2
  \int_0^{\ell_2} d\ell_3 \int_0^{\ell_3} d\ell_4 \notag
  \\[0.2cm]
  &\quad\,
  \text{tr}_{(NZ)} \bigg[ e^{-\ell K^2} \Big\{ 8 F^\mu{}_\nu(\ell_1) F^\nu{}_\rho(\ell_2) F^\rho{}_\sigma(\ell_3) F^\sigma{}_\mu(\ell_4) 
  + 16 F_{\mu\nu}(\ell_1) F^{\mu\rho}(\ell_2) F^{\nu\sigma}(\ell_3) F_{\rho\sigma}(\ell_4) \notag
  \\
  &\qquad
  - 4 F_{\mu\nu}(\ell_1) F^{\mu\nu}(\ell_2) F_{\rho\sigma}(\ell_3) F^{\rho\sigma}(\ell_4)
  - 2 F_{\mu\nu}(\ell_1) F_{\rho\sigma}(\ell_2) F^{\mu\nu}(\ell_3) F^{\rho\sigma}(\ell_4) \Big\} \bigg], \label{eq:Veff2}
  \\[0.2cm]
  &\quad\,
  F_{\mu\nu}(\ell) \equiv e^{\ell K^2} F_{\mu\nu} \, e^{-\ell K^2}. \notag
\end{alignat}
where the trace is taken for $(NZ)\times (NZ)$ matrix.
Furthermore the $(NZ) \times (NZ)$ matrix is decomposed into the product of $N \times N$ matrix and $Z \times Z$ matrix.
Indeed, $e^{-\ell K^2} = e^{-\ell Q^2} \otimes e^{-\ell P^2}$,
and non-zero component of the field strength is $F_{\alpha\beta}(\ell) = G_{\alpha\beta}(\ell) \otimes {\bf 1}_Z$ with
\begin{alignat}{3}
  &G_{\alpha\beta}(\ell) \equiv e^{\ell Q^2} G_{\alpha\beta} \, e^{-\ell Q^2}, \qquad
  G_{\alpha\beta} = \begin{pmatrix} 0 & \dot{Q}_b \\ - \dot{Q}_a & -i[Q_a,Q_b] \end{pmatrix}, \label{eq:G}
\end{alignat}
where $\alpha,\beta = \tau,1,2,3$. $Q_a$ is defined in the eq.~(\ref{eq:Q_a}), 
and we set $\dot{\tilde{x}}_a = 0$ for slowly moving test D0-brane.
Finally the effective potential (\ref{eq:Veff2}) is expressed as
\begin{alignat}{3}
  V_{\text{eff}} \big|_{M^4} &= 
  - \frac{1}{2\sqrt{\pi}} \int_0^\infty \frac{d\ell}{\ell^{3/2}} \int_0^\ell d\ell_1 \int_0^{\ell_1} d\ell_2
  \int_0^{\ell_2} d\ell_3 \int_0^{\ell_3} d\ell_4 \, \text{tr}_Z \Big( e^{-\ell P^2} \Big) \notag
  \\[0.2cm]
  &\quad\,
  \text{tr}_N \bigg[ e^{-\ell Q^2} \Big\{ 8 G^\alpha{}_\beta(\ell_1) G^\beta{}_\gamma(\ell_2) G^\gamma{}_\delta(\ell_3) G^\delta{}_\alpha(\ell_4) 
  + 16 G_{\alpha\beta}(\ell_1) G^{\alpha\gamma}(\ell_2) G^{\beta\delta}(\ell_3) G_{\gamma\delta}(\ell_4) \notag
  \\
  &\qquad
  - 4 G_{\alpha\beta}(\ell_1) G^{\alpha\beta}(\ell_2) G_{\gamma\delta}(\ell_3) G^{\gamma\delta}(\ell_4)
  - 2 G_{\alpha\beta}(\ell_1) G_{\gamma\delta}(\ell_2) G^{\alpha\beta}(\ell_3) G^{\gamma\delta}(\ell_4) \Big\} \bigg]. \label{eq:Veff3}
\end{alignat}
Thus $\text{tr}_{(NZ)}$ is factorized into $\text{tr}_Z$ and $\text{tr}_N$.
If we take the large $Z$ limit with $R_u$ fixed, the trace $\text{tr}_Z$ is transformed into Gaussian integral.
\begin{alignat}{3}
  &\text{tr}_Z \Big( e^{-\ell P^2} \Big) = \sum_{n_4=1}^{Z_4} \cdots \sum_{n_9=1}^{Z_9} e^{- \ell p_u^2} \sim
  \frac{\pi^3 Z}{2^6 M_6} \frac{1}{\ell^3}, \label{eq:trZ}
\end{alignat}
where $p_u = 2\pi \tilde{R}_u n_u/Z_u$, $\tilde{R}_u = R_u /(2\pi\ell_s^2)$
and $M_6 = \prod_{u=4}^9 2\pi \tilde{R}_u$.
Note that the dependence on $\tilde{x}_a$ appears through $Q^2$ in the eq.~(\ref{eq:Veff3}).


\section{Effective Potentials via Smeared Fuzzy Objects}


Let us calculate the effective potentials (\ref{eq:Veff3}) between the smeared fuzzy objects and the test D0-brane
by using eqs.~(\ref{eq:Q_a}), (\ref{eq:G}) and (\ref{eq:trZ}).
We assume that the test D0-brane is located on the $x_3$ axis without loss of generality.
So we set $\tilde{x} = \tilde{x}_3$ in this section.


\subsection{Effective Potential via Smeared Fuzzy Sphere}


In this subsection, we calculate the effective potential between the smeared fuzzy sphere and the test D0-brane.
Then $Q^2$ in eq.~(\ref{eq:Q_a}) is evaluated as
\begin{alignat}{3}
  &Q^2 = \Big( \frac{N^2-1}{4} \tilde{r}^2 + \tilde{x}^2 \Big) {\bf 1}_N - \tilde{r} \tilde{x} \Sigma_3, 
\end{alignat}
and the field strength $G_{\alpha\beta}(\ell)$ is given by
\begin{alignat}{3}
  &G_{\alpha\beta}(\ell) = 
  \begin{pmatrix} 
    0 & \dot{\tilde{r}} \frac{\Sigma_b(\theta)}{2} \\[0.1cm]
    - \dot{\tilde{r}} \frac{\Sigma_a(\theta)}{2} & \tilde{r}^2 \epsilon_{abc} \frac{\Sigma^c(\theta)}{2}
  \end{pmatrix}.
\end{alignat}
Here we defined $\theta \equiv 2 \ell \tilde{r} \tilde{x}$ and
$\Sigma_a(\theta) \equiv e^{\ell Q^2} \Sigma_a e^{-\ell Q^2}$. 
And the explicit form of $\Sigma_a(\theta)$ is given as follows.
\begin{alignat}{3}
  &\begin{pmatrix} \Sigma_1(\theta) \\ \Sigma_2(\theta) \\ \Sigma_3(\theta) \end{pmatrix}
  = \begin{pmatrix} \cosh \theta & -i \sinh \theta & 0 \\ i \sinh \theta & \cosh \theta & 0 \\ 0 & 0 & 1 \end{pmatrix}
  \begin{pmatrix} \Sigma_1 \\ \Sigma_2 \\ \Sigma_3 \end{pmatrix}. 
\end{alignat}
By inserting the above expressions into eq.~(\ref{eq:Veff3}), the effective potential at $M^4$ order is evaluated as follows.
\begin{alignat}{3}
  V_{\text{eff}} \big|_{M^4} &= - \frac{1}{2^8\sqrt{\pi}} 
  \frac{\pi^3 Z}{M_6} \frac{1}{(2\tilde{r}\tilde{x})^{1/2}}
  \int_0^\infty \!\!\! \frac{d\theta}{\theta^{9/2}} \int_0^\theta \!\!\! d\theta_1 \int_0^{\theta_1} \!\!\! d\theta_2
  \int_0^{\theta_2} \!\!\! d\theta_3 \int_0^{\theta_3} \!\!\! d\theta_4 \notag
  \\
  &\quad\;
  e^{-\big( \frac{N^2-1}{4} \tilde{r}^2 + \tilde{x}^2 \big) \frac{\theta}{2\tilde{r}\tilde{x}}} \,
  \text{tr}_N \Big[ e^{\theta \frac{\Sigma_3}{2}} \Big\{ (\dot{\tilde{r}}^2 - \tilde{r}^4)^2 
  \big( \Sigma_a(\theta_1) \Sigma^a(\theta_2) \Sigma_b(\theta_3) \Sigma^b(\theta_4) \notag
  \\
  &\quad\,
  + \Sigma_a(\theta_1) \Sigma_b(\theta_2) \Sigma^b(\theta_3) \Sigma^a(\theta_4)
  + \Sigma_a(\theta_1) \Sigma_b(\theta_2) \Sigma^a(\theta_3) \Sigma^b(\theta_4) \big) \notag
  \\
  &\quad\,
  + 4 \dot{\tilde{r}}^2 \tilde{r}^4 \big( \Sigma_a(\theta_1) \Sigma^a(\theta_2) \Sigma_b(\theta_3) \Sigma^b(\theta_4)
  - \Sigma_a(\theta_1) \Sigma_b(\theta_2) \Sigma^b(\theta_3) \Sigma^a(\theta_4) \big) \Big\} \Big] \notag
  \\
  &= - \frac{1}{2^8\sqrt{\pi}} \frac{\pi^3 Z}{M_6} \frac{1}{(2\tilde{r}\tilde{x})^{1/2}}
  \int_0^\infty \frac{d\theta}{\theta^{9/2}} \notag
  \\
  &\quad\;
  e^{-\big( \frac{N^2-1}{4} \tilde{r}^2 + \tilde{x}^2 \big)\frac{\theta}{2\tilde{r}\tilde{x}}} \,
  \text{tr}_N \Big[ e^{\theta \frac{\Sigma_3}{2}} \Big\{ c_1^8 J_1(\theta)
  + 4 \dot{\tilde{r}}^2 \tilde{r}^4 J_2(\theta) \Big\} \Big]. \label{eq:Veffsph}
\end{alignat}
Here we used the energy conservation (\ref{eq:Esph}),
which is written as $- \dot{\tilde{r}}^2 + \tilde{r}^4 = c_1^4$ for Euclidean time.
Therefore the coefficient of $J_1(\theta)$ term in the trace is time independent.
$J_1(\theta)$ and $J_2(\theta)$ are power series of $\Sigma_3^n$ 
and their explicit forms are given in the appendix~\ref{sec:J12}.
Now we take the large $N$ limit by keeping the size of the fuzzy sphere. 
Then the trace is transformed into integral.
\begin{alignat}{3}
  &\text{tr}\Big( e^{\theta \frac{\Sigma_3}{2}} \Sigma_3^n \Big) = 
  2^n \frac{d^n}{d\theta^n} \text{tr}\Big( e^{\theta \frac{\Sigma_3}{2}} \Big)
  \sim 2^{n+1} \frac{d^n}{d\theta^n} \Big( \frac{\sinh(\frac{N\theta}{2})}{\theta} \Big). \label{eq:trN}
\end{alignat}
Here the representation of $\Sigma_3$ was chosen as $(\Sigma_3)_{m,n} = (N-2m+1) \delta_{m,n}$.
Finally, by inserting the above equation into the eq.~(\ref{eq:Veffsph}),
we obtain the effective potential at $M^4$ order as
\begin{alignat}{3}
  V_{\text{eff}} \big|_{M^4} 
  &= - \frac{\pi^3 N Z}{2^{11} M_6} \frac{N^4 c_1^8}{(2\tilde{R}_\text{sph} \tilde{x})^{\frac{1}{2}}} 
  \bigg\{ \sqrt{\frac{\tilde{R}_\text{sph} }{2\tilde{x}} \!+\! \frac{\tilde{x}}{2\tilde{R}_\text{sph}} \!+\! 1} 
  - \sqrt{\frac{\tilde{R}_\text{sph}}{2\tilde{x}} \!+\! \frac{\tilde{x}}{2\tilde{R}_\text{sph}} \!-\! 1} 
  + \mathcal{O} \Big(\frac{1}{N^2}\Big) \bigg\}, \label{eq:Veffsph2}
\end{alignat}
where $\tilde{R}_\text{sph} \sim \tilde{r}N/2$ is typical mass scale of the fuzzy sphere.
The final expression is derived by employing Mathematica.
Notice that the integral of the $J_2$ part behaves like $\mathcal{O}(1/N^2)$ in the above,
so it is neglected in the large $N$ limit. See appendix~\ref{sec:J12} for details.
Although $\tilde{R}_\text{sph}$ is time dependent, we assume that the fuzzy sphere is oscillating slowly around $t=0$
and $\tilde{R}_\text{sph}$ is finite.
Then if we take $\tilde{R}_\text{sph} \ll \tilde{x}$, where the test D0-brane is far from the fuzzy sphere in 3 dimensions, 
the above effective potential becomes
\begin{alignat}{3}
  V_{\text{eff}} \big|_{M^4} 
  &\sim - \frac{\pi^3 N Z}{2^{11} M_6} \frac{N^4 c_1^8}{\tilde{x}} 
  \sim - \frac{\pi^3 (NZ)}{2^5 M_6} \Big( \frac{E_\text{sph} \lambda}{(NZ)^2} \Big)^2 \frac{1}{\tilde{x}}. \label{eq:Veffsph3}
\end{alignat}
In the above, the eq.~(\ref{eq:Esphcyl}) is used.
This result should be compared with that of the near horizon geometry of the smeared black 0-brane.


\subsection{Effective Potential via Smeared Fuzzy Cylinder}


In this subsection, we analyze the effective potential between the fuzzy cylinder and the test D0-brane.
The $Q^2$ in the eq.~(\ref{eq:Q_a}) is evaluated as
\begin{alignat}{3}
  &Q^2 = \big( \tilde{\rho}^2 + \tilde{x}^2 \big) {\bf 1}_N + \tilde{l}^2 \Xi_3^2  - 2 \tilde{l} \tilde{x} \Xi_3,
\end{alignat}
and the field strength $G_{\alpha\beta}(\ell)$ is given by
\begin{alignat}{3}
  &G_{\alpha\beta}(\ell) = 
  \begin{pmatrix} 
    0 & \dot{\tilde{\rho}} \, \Xi_b(\ell) & 0 \\[0.1cm]
    - \dot{\tilde{\rho}} \, \Xi_a(\ell) & 0 & - \tilde{l} \tilde{\rho} \, \epsilon_{ac} \Xi^c(\ell) \\[0.1cm]
    0 & \tilde{l} \tilde{\rho} \, \epsilon_{bc} \Xi^c(\ell) & 0 
  \end{pmatrix},
\end{alignat}
where $a,b=1,2$ and $\epsilon_{ab}$ is an antisymmetric tensor. 
$\Xi_a(\ell) \equiv e^{\ell Q^2} \Xi_a e^{-\ell Q^2}$ is explicitly evaluated as
\begin{alignat}{3}
  &\big( \Xi_1(\ell) \big)_{mn} = \frac{1}{2} e^{\ell \lambda_m} \delta_{m+1,n} + \frac{1}{2} e^{- \ell \lambda_n} \delta_{m,n+1}, \notag
  \\
  &\big( \Xi_2(\ell) \big)_{mn} = - \frac{i}{2} e^{\ell \lambda_m} \delta_{m+1,n} + \frac{i}{2} e^{- \ell \lambda_n} \delta_{m,n+1}, 
  \\
  &\,\lambda_m \equiv - 2 \tilde{l}^2 \Big( m + \frac{\tilde{x}}{\tilde{l}} \Big) - \tilde{l}^2.\notag
\end{alignat}
By inserting the above expressions into eq.~(\ref{eq:Veff3}), the effective potential at $M^4$ order is evaluated as follows.
\begin{alignat}{3}
  V_{\text{eff}} \big|_{M^4} &= - \frac{1}{2^4\sqrt{\pi}} \frac{\pi^3 Z}{M_6}
  \int_0^\infty \!\!\! \frac{d\ell}{\ell^{9/2}} \int_0^\ell \!\!\! d\ell_1 \int_0^{\ell_1} \!\!\! d\ell_2
  \int_0^{\ell_2} \!\!\! d\ell_3 \int_0^{\ell_3} \!\!\! d\ell_4 \notag
  \\
  &\quad\;
  e^{- \ell ( \tilde{\rho}^2 + \tilde{x}^2)} \,
  \text{tr} \Big[ e^{- \ell \big( \tilde{l}^2 \Xi_3^2 - 2 \tilde{l} \tilde{x} \Xi_3 \big)} 
  \Big\{ (\dot{\tilde{\rho}}^2 - \tilde{l}^2 \tilde{\rho}^2)^2 
  \big( \Xi_a(\ell_1) \Xi^a(\ell_2) \Xi_b(\ell_3) \Xi^b(\ell_4) \notag
  \\
  &\quad\,
  + \Xi_a(\ell_1) \Xi_b(\ell_2) \Xi^b(\ell_3) \Xi^a(\ell_4)
  + \Xi_a(\ell_1) \Xi_b(\ell_2) \Xi^a(\ell_3) \Xi^b(\ell_4) \big) \notag
  \\
  &\quad\,
  + 4 \tilde{l}^2 \tilde{\rho}^2 \dot{\tilde{\rho}}^2 \big( \Xi_a(\ell_1) \Xi^a(\ell_2) \Xi_b(\ell_3) \Xi^b(\ell_4)
  - \Xi_a(\ell_1) \Xi_b(\ell_2) \Xi^b(\ell_3) \Xi^a(\ell_4) \big) \Big\} \Big] \notag
  \\
  &= - \frac{1}{2^4\sqrt{\pi}} \frac{\pi^3 Z}{M_6} \int_0^\infty \!\!\! \frac{d\ell}{\ell^{9/2}} 
  e^{- \ell \tilde{\rho}^2 } \,
  \text{tr}_N \Big[ e^{- \ell \tilde{l}^2 \big( \Xi_3 - \frac{\tilde{x}}{\tilde{l}} {\bf 1} \big)^2 } \,
  \Big\{ \tilde{l}^4 c_3^4 L_1(\ell) + 4 \tilde{l}^2 \tilde{\rho}^2 \dot{\tilde{\rho}}^2 L_2(\ell) \Big\} \Big]. \label{eq:Veffcyl}
\end{alignat}
Here we used the energy conservation (\ref{eq:Ecyl}) for the Euclidean time, that is,
$- \dot{\tilde{\rho}}^2 + \tilde{l}^2 \tilde{\rho}^2 = \tilde{l}^2 c_3^2$.
Due to this relation, $L_1(\ell)$ part in the trace is time independent.
$L_1(\ell)$ and $L_2(\ell)$ are diagonal matrices and their explicit forms are given in the appendix~\ref{sec:L12}.
Now we take the density of D0-branes per length infinite. 
Then the trace is transformed into integral.
\begin{alignat}{3}
  &\text{tr}\Big( e^{- \ell \tilde{l}^2 \big( \Xi_3 - \frac{\tilde{x}}{\tilde{l}} {\bf 1} \big)^2 } F \Big) 
  \sim \int_{-\infty}^\infty d\zeta \, e^{-\ell \tilde{l}^2 \zeta^2} f ( - 2 \tilde{l}^2 \zeta - \tilde{l}^2 ). \label{eq:tr2}
\end{alignat}
Here $F$ is some diagonal matrix whose component is given by $F_{mn} = f(\lambda_m) \delta_{m,n}$.
Since the length of the fuzzy cylinder is infinite, the range of the integral also becomes infinite.
Then the $\tilde{x}$ dependence disappears by shifting the origin.
Finally, by inserting the above equation into eq.~(\ref{eq:Veffcyl}), 
the effective potential at $M^4$ order becomes
\begin{alignat}{3}
  V_{\text{eff}} \big|_{M^4} &\sim - \frac{\pi^3 Z}{2^4 M_6} \int_{-\infty}^\infty d\zeta 
  \Big\{ \tilde{l}^4 c_3^4 \, \hat{L}_1 \big( \zeta,\tfrac{\tilde{\rho}}{\tilde{l}} \big)
  + \tilde{l}^2 \tilde{\rho}^2 \dot{\tilde{\rho}}^2 \, \hat{L}_2 \big( \zeta,\tfrac{\tilde{\rho}}{\tilde{l}} \big) 
  \Big\}. \label{eq:Veffcyl2}
\end{alignat}
The last expression is derived by employing Mathematica.
The functions $\hat{L}_1$ and $\hat{L}_2$ depend on $\zeta$ and $\tilde{\rho}/\tilde{l}$, and
the explicit forms are given in the appendix~\ref{sec:L12}.
The effective potential of eq.~(\ref{eq:Veffcyl2}) shows that 
there is no force between the fuzzy cylinder and the test D0-brane at $M^4$ order.


\section{Comparison with the Gravity Side}


In this section we review the properties of the smeared black 0-brane and compare the effective potentials
for the test D0-brane with those of the previous section.
The black 0-brane solution is obtained by boosting the 11 dimensional black hole along 11th direction.
In a similar way, the smeared black 0-brane solution can be constructed
by boosting the smeared black hole along the 11th direction\cite{Hyakutake:2016sig}.
The metric, the dilaton field and the R-R 1-form field for the smeared black 0-brane are written as
\begin{alignat}{3}
  &ds_{10}^2 = - H^{-\frac{1}{2}} F dt^2 + H^{\frac{1}{2}} \big( F^{-1} dr^2 
  + r^2 d\Omega_2^2 + dx_u^2 \big), \notag
  \\
  &e^\phi = H^\frac{3}{4}, \qquad C^{(1)} = \sqrt{1+\alpha} \, (1 - H^{-1}) dt, \label{eq:smearD0}
  \\
  &H = 1 + \frac{r_-}{r}, \qquad F = 1 - \frac{r_- \alpha}{r}. \notag
\end{alignat}
Here $x_u \,(u=4,\cdots,9)$ labels the smeared directions.
The solution has two parameters $r_-$ and $\alpha$, and the latter corresponds to the boost parameter.

Let us evaluate physical quantities of the smeared black 0-brane.
The event horizon is located at $r_\text{h} = r_- \alpha$, and the temperature $T$ and
electric potential $\Phi$ are given by
\begin{alignat}{3}
  T &= \frac{1}{4\pi} H^{-1/2} \frac{dF}{dr} \Big|_{r_\text{h}} 
  = \frac{1}{4\pi r_- \alpha} \sqrt{\frac{\alpha}{1+\alpha}}, \label{eq:Tsmear}
  \\
  \Phi &= C^{(1)}_t \Big|_{r_\text{h}} = \frac{1}{\sqrt{1+\alpha}}. \notag
\end{alignat}
The ADM mass $M$ and the R-R charge $Q$ of the smeared black 0-brane are evaluated as usual,
and the results become
\begin{alignat}{3}
  M &= \frac{4\pi V_6}{2 \kappa_{10}^2} r_- \alpha
  \Big( 2 + \frac{1}{\alpha} \Big), \qquad
  Q = \frac{4\pi V_6}{2 \kappa_{10}^2} (\sqrt{1 + \alpha}) r_-.
\end{alignat}
$V_6 = \prod_{u=4}^9 2\pi R_u$ is the volume of the compactified 6 directions and $2\kappa_{10}^2 = (2\pi)^7 \ell_s^8 g_s^2$
is the 10 dimensional gravitational constant. 
$\ell_s$ is the string length and $g_s$ is the string coupling constant.
The extremal limit corresponds to $\alpha \to 0$.

Next let us consider the near horizon limit of the smeared black 0-brane.
The near horizon limit is defined so that physical quantities of 
the dual gauge theory become finite\cite{Maldacena:1997re}.
Thus the near horizon limit for the black 0-brane is defined as\cite{Itzhaki:1998dd}
\begin{alignat}{3}
  r \to 0 \quad \text{with $\;U = \frac{r}{\ell_s^2}\;$ and $\;\lambda = \frac{g_s N'}{(2\pi)^2 \ell_s^3}\;$ fixed.}
\end{alignat}
Here $U$ is a typical energy scale of the system. 
The 't Hooft coupling is denoted by $\lambda = g_\text{YM}^2 N'$ and $N'=NZ$ is the number of the smeared D0-branes.
Note that the energy scale at the horizon $U_\text{h} = \frac{r_- \alpha}{\ell_s^2}$ is also fixed.
In terms of $\alpha$ and $r_-$, the near horizon limit is defined as
\begin{alignat}{3}
  \alpha \to 0 \quad \text{with $\;\frac{r}{r_- \alpha}\;$ and $\;\frac{r_- \alpha}{\ell_s^2}\;$ fixed.} 
  \label{eq:lim}
\end{alignat}
Let us examine $\alpha \to 0$ limit more carefully.
Since the black 0-brane corresponds to the D0-brane, the R-R charge of the D0-branes should be
\begin{alignat}{3}
  Q = \frac{N'}{\ell_s g_s}.
\end{alignat}
Furthermore, since the black 0-brane is smeared into 6 spatial directions, we should fix 
typical mass scale for the compactified 6 spatial directions. 
Namely we fix $M_6 = \prod_{u=4}^9 2\pi \tilde{R}_u$.
Then, in the near horizon limit, $\alpha$ goes to zero like
\begin{alignat}{3}
  \alpha \;\to\; \frac{M_6 U_\text{h}}{2\pi^2 \lambda} \ell_s^4.
\end{alignat}
Note that $r_-$ goes to the infinity through the relation $r_- = U_\text{h} \ell_s^2/\alpha$,
and $H$ and $F$ in eq.~(\ref{eq:smearD0}) are written as
\begin{alignat}{3}
  H \to \frac{1}{\alpha} \frac{U_\text{h}}{U}, \qquad F = 1 - \frac{U_\text{h}}{U}.
\end{alignat}

Thermodynamics of the near horizon geometry of the smeared black 0-brane becomes as follows.
The temperature in (\ref{eq:Tsmear}) becomes
\begin{alignat}{3}
  T &= \frac{M_6^{1/2}}{4\sqrt{2}\pi^2 \lambda^{1/2} U_\text{h}^{1/2}}, \label{eq:Tnear}
\end{alignat}
and the internal energy $E = M-Q$ is expressed as
\begin{alignat}{3}
  \frac{E}{N'^2} &= \frac{3M_6 U_\text{h}}{16\pi^4 \lambda^2}
  = \frac{3M_6^2}{2(2\pi)^8 \lambda^3 T^2}. \label{eq:ESnear}
\end{alignat}

Finally we examine a test D0-brane moving around the smeared black 0-brane.
Let us consider the potential energy for the test D0-brane,
which is moving only along the radial direction.
With this assumption, the Lagrangian for the D0-brane in the background of 
the smeared black 0-brane (\ref{eq:smearD0}) becomes
\begin{alignat}{3}
  \mathcal{L} &= - T_0 e^{-\phi} \sqrt{- g_{\mu\nu} \dot{x}^\mu \dot{x}^\nu} - T_0 C^{(1)}_t \notag
  \\
  &= - T_0 e^{-\phi} H^{-\frac{1}{4}} F^{\frac{1}{2}} \sqrt{1 - H F^{-2} \dot{r}^2}
  - T_0 \sqrt{1+\alpha} (1-H^{-1}).
\end{alignat}
And the momentum conjugate to $r$ is defined as
\begin{alignat}{3}
  p_r = \frac{\pa \mathcal{L}}{\pa \dot{r}} = T_0 H^{-1} F^{\frac{1}{2}} 
  \frac{H F^{-2} \dot{r}}{\sqrt{1 - H F^{-2} \dot{r}^2}}.
\end{alignat}
By using the above equation $\dot{r}$ is expressed in terms of $p_r$, 
and the Hamiltonian of the D0-brane is evaluated as
\begin{alignat}{3}
  \mathcal{H} &= H^{-1} F^{\frac{1}{2}} \sqrt{T_0^2 + H F p_r^2}
  + T_0 \sqrt{1+\alpha} (1-H^{-1}).
\end{alignat}
If the momentum is small enough, we can expand the above with respect to the momentum 
and read off the potential energy as
\begin{alignat}{3}
  V &= T_0 H^{-1} F^{\frac{1}{2}} + T_0 \sqrt{1+\alpha} (1-H^{-1}).
\end{alignat}
The first term corresponds to the attractive force by the gravity and the second term
does to the repulsive force due to the R-R background.
In the classical (or $1 \ll r$) and near horizon limits, the potential becomes
\begin{alignat}{3}
  V - T_0 &\;\sim\; T_0 \alpha \frac{U}{U_\text{h}} (\sqrt{F} - 1) + T_0 \frac{\alpha}{2} \notag
  \\
  &\;\sim\; - \frac{M_6 U_\text{h}^2 N'}{64\pi^4 \lambda^2} \frac{1}{U} \notag
  \\
  &\;=\; - \frac{4 \pi^4 (NZ)}{M_6} \Big( \frac{E \lambda}{3 (NZ)^2} \Big)^2 \frac{1}{U} \notag
  \\
  &\;=\; - \frac{2 \pi^3 (NZ)}{9 M_6} \Big( \frac{E \lambda}{(NZ)^2} \Big)^2 \frac{1}{\tilde{x}}. \label{eq:Veffgra}
\end{alignat}
The rest mass of the D0-brane is subtracted in the above, since it is divergent constant 
in the near horizon limit.
In the last line, we used eq.~(\ref{eq:ESnear}), $N'=NZ$ and $U = x/\ell_s^2 = 2\pi \tilde{x}$.
The qualitative feature of the eq.~(\ref{eq:Veffgra}) surely matches with the eq.~(\ref{eq:Veffsph3}).

So far we have smeared the 4 dimensional black hole along 6 spatial directions,
and boosted it along the 11th direction. And the solution is given by eq.~(\ref{eq:smearD0}).
Then we might try to smear the 3 dimensional black hole along 7 spatial directions,
and boost it along the 11th direction. However, there is no 3 dimensional black hole
which is asymptotic to the flat spacetime\cite{Ida:2000jh}. This means that there is no black 0-brane
which is smeared along 7 spatial directions. So the effective potential between the 
black 0-brane and test D0-brane should be trivial.
This is consistent with the result (\ref{eq:Veffcyl2}), which does not depend on $\tilde{x}$.


\section{Conclusion and Discussion}


In this paper, we proposed that the fuzzy configurations of D0-branes in the BFSS matrix model
would correspond to the microstates of the smeared black 0-brane in the near horizon limit.
The fuzzy configurations are constructed by smearing the fuzzy objects in 3 dimensions 
into 6 spatial directions.
Since the fuzzy objects have the internal energy compared with the static case, 
they are time dependent and non-BPS states.
Thus the fuzzy configurations would correspond to the microstates of the non-extremal black 0-brane
in the near horizon limit.
As a non-trivial check, we evaluated the one-loop effective potential for the test D0-brane
in the background of the smeared fuzzy sphere. 
We found that the effective potential for the test D0-brane behaves like the eq.~(\ref{eq:Veffsph3})
in the BFSS matrix model.
On the other hand, the effective potential was also evaluated from the gravity side like the eq.~(\ref{eq:Veffgra}).
These two results match up to the numerical factor, so this shows an evidence that the smeared fuzzy objects
are  the microstates of the black hole.
Furthermore, we also evaluated the one-loop effective potential for the test D0-brane 
in the background of the smeared fuzzy cylinder.
In this case, the effective potential becomes trivial, and it agrees with the fact that
there is no asymptotically flat black hole in 3 dimensions.

Although the qualitative features of the smeared fuzzy objects match with those of the smeared black 0-brane 
in the gravity side, we still have the discrepancy in the numerical coefficients.
This is similar to the case of non-extremal black 3-brane thermodynamics\cite{Gubser:1998nz}.
In order to cure this problem from the gravity side, we need to take into account $\alpha'$ corrections in 
type IIA superstring theory. This will modify the form of $F(r)$ in the metric (\ref{eq:smearD0}) and 
the mass of the black 0-brane will be renormalized as argued in ref.~\cite{Hyakutake:2013vwa}. 

In this paper, we focused on irreducible representations in the eq.~(\ref{eq:fuzzy}).
It is possible, however, to consider reducible ones which correspond to multi fuzzy objects.
For example, we divide the size of the matrix $N$ for the fuzzy object into $n$ pieces like $N = \sum_{i=1}^n N_i$, and
prepare parameters $d_i\, (i=1,\cdots,n)$ so as to satisfy $N^3 c_1^4 = \sum_{i=1}^n N_i^3 d_i^4$.
Then we construct the fuzzy object out of $n$ fuzzy spheres, each of which has the the matrix size $N_i$ and
the internal energy $E_\text{sph,$i$} = N_i^3 d_i^4/(8 g_\text{YM}^2)$.
This fuzzy object has the same internal energy as the eq.~(\ref{eq:Esph}) in the large $N_i$ limit.
And the effective potential for the test D0-brane (\ref{eq:Veffsph3}) is modified as follows.
\begin{alignat}{3}
  V_{\text{eff}} \big|_{M^4,J_1} 
  &\sim - \frac{\pi^3 Z}{2^{11} M_6} \frac{\sum_{i=1}^n N_i^5 d_i^8}{\tilde{x}}. \label{eq:Veffsph4}
\end{alignat}
The numerical coefficient is different from the eq.~(\ref{eq:Veffsph3}), but the order is almost the same.
For instance, if we choose $N_i/N \sim 1/n$ and $(N_i/N)^3 (d_i/c_1)^4 \sim 1/n$ for all $i$,
we obtain $\sum_{i=1}^n (N_i/N)^5 (d_i/c_1)^8 \sim 1$.
Since these configurations give the same internal energy as the single fuzzy sphere,
these will be the microstates of the smeared black 0-brane in the near horizon limit.
Note that this proposal is similar to the notion of fuzz ball for the black hole\cite{Mathur:2005zp}.

In this paper, we considered the fuzzy objects which have axial symmetry.
It is possible to relax this ansatz to construct generic configuration\cite{Shimada:2003ks},
and it will also contribute to the microstates of the black hole.
For future directions, it is interesting to examine the multi-shell model which is proposed
as an alternative black hole evaporation mechanism in ref.~\cite{Kawai:2013mda,Ho:2015vga}.
The similar situation can be analyzed by using the multi fuzzy spheres discussed in the above.
Notice that the time evolutions of the fuzzy configurations are quite complicated even for the two D0-branes case. (See fig.~\ref{fig:czN=2}.)
This shows that black hole has a chaotic behavior as recently studied in refs.~\cite{Sekino:2008he}-\cite{Berkowitz:2016znt}.
Since the fuzzy configurations are time dependent, it is also interesting to deal with out-of-equilibrium properties of those
in the BFSS matrix model\cite{Craps:2016cgo}.

\section*{Acknowledgement}

The author would like to thank NTU, NCTS and DIAS for their warm hospitality.
Especially the author would like to thank Yuhma Asano, Masanori Hanada, Pei-Ming Ho, Denjoe O'Connor,
Tadashi Okazaki.
This work was partially supported by the Ministry of Education, Science, 
Sports and Culture, Grant-in-Aid for Scientific Research (C) 17K05405, 2017.

\appendix


\section{Calculation of $J_1$ and $J_2$} \label{sec:J12}


Definition of $J_1(\theta)$ and explicit expression are given as follows.
\begin{alignat}{3}
  J_1(\theta) &\equiv \int_0^\theta \!\!\! d\theta_1 \int_0^{\theta_1} \!\!\! d\theta_2
  \int_0^{\theta_2} \!\!\! d\theta_3 \int_0^{\theta_3} \!\!\! d\theta_4 
  \Big( \Sigma_a(\theta_1) \Sigma^a(\theta_2) \Sigma_b(\theta_3) \Sigma^b(\theta_4) \notag
  \\[-0.1cm]
  &\quad\,
  + \Sigma_a(\theta_1) \Sigma_b(\theta_2) \Sigma^b(\theta_3) \Sigma^a(\theta_4)
  + \Sigma_a(\theta_1) \Sigma_b(\theta_2) \Sigma^a(\theta_3) \Sigma^b(\theta_4) \Big) \notag
  \\
  &= \int_0^\theta \!\!\! d\theta_1 \int_0^{\theta_1} \!\!\! d\theta_2
  \int_0^{\theta_2} \!\!\! d\theta_3 \int_0^{\theta_3} \!\!\! d\theta_4 \Big( 
  - 4 \sinh (\theta_1 \!- \! \theta_2 \!+\! \theta_3 \!-\! \theta_4) \big( (N^2 \!-\! 1) \Sigma_3 \!-\! \Sigma_3^3 \big) \notag
  \\
  &\quad\,
  - 8 \sinh (\theta_1 \!+\! \theta_2 \!-\! \theta_3 \!-\! \theta_4) \big( (N^2 \!-\! 3) \Sigma_3 \!-\! \Sigma_3^3 \big) \notag
  \\
  &\quad\,
  + \cosh (\theta_1 \!+\! \theta_2 \!-\! \theta_3 \!-\! \theta_4) 
  \big( (N^2 \!-\! 9)(N^2 \!-\! 1) \!-\! 2 (N^2 \!-\! 11) \Sigma_3^2 \!+\! \Sigma_3^4 \big) \notag
  \\
  &\quad\,
  + \cosh (\theta_1 \!-\! \theta_2 \!+\! \theta_3 \!-\! \theta_4) 
  \big( (N^2 \!-\! 1)^2 \!-\! 2 (N^2 \!-\! 3) \Sigma_3^2 \!+\! \Sigma_3^4 \big) \notag
  \\
  &\quad\, 
  + \cosh (\theta_1 \!-\! \theta_2 \!-\! \theta_3 \!+\! \theta_4) 
  \big( (N^2 \!-\! 1)^2 \!-\! 2 (N^2 \!+\! 1) \Sigma_3^2 \!+\! \Sigma_3^4 \big)
  - 2 \sinh (\theta_1 \!-\! \theta_2) \Sigma_3^3 \notag
  \\
  &\quad\,
  + 2 \sinh (\theta_1 \!-\! \theta_3) \big( (N^2 \!-\! 1) \Sigma_3 \!-\! 2 \Sigma_3^3 \big) 
  + 2 \sinh (\theta_1 \!-\! \theta_4) \big( 2 (N^2 \!-\! 3) \Sigma_3 \!-\! 3 \Sigma_3^3 \big)  \notag
  \\
  &\quad\,
  - 2 \sinh (\theta_2 \!-\! \theta_3) \Sigma_3^3
  + 2 \sinh (\theta_2 \!-\! \theta_4) \big( (N^2 \!-\! 1) \Sigma_3 \!-\! 2 \Sigma_3^3 \big)
  - 2 \sinh (\theta_3 \!-\! \theta_4) \Sigma_3^3 \notag
  \\
  &\quad\,
  + \cosh (\theta_1 \!-\! \theta_2) \big( (N^2 \!-\! 1) \Sigma_3^2 \!-\! \Sigma_3^4 \big)
  + \cosh (\theta_1 \!-\! \theta_3) \big( (N^2 \!-\! 5) \Sigma_3^2 \!-\! \Sigma_3^4 \big) \notag
  \\
  &\quad\,
  + \cosh (\theta_1 \!-\! \theta_4) \big( 4 (N^2 \!-\! 1) + (N^2 \!-\! 13) \Sigma_3^2 \!-\! \Sigma_3^4 \big)
  + \cosh (\theta_2 \!-\! \theta_3) \big( (N^2 \!-\! 1) \Sigma_3^2 \!-\! \Sigma_3^4 \big) \notag
  \\
  &\quad\,
  + \cosh (\theta_2 \!-\! \theta_4) \big( (N^2 \!-\! 5) \Sigma_3^2 \!-\! \Sigma_3^4 \big)
  + \cosh (\theta_3 \!-\! \theta_4) \big( (N^2 \!-\! 1) \Sigma_3^2 \!-\! \Sigma_3^4 \big) 
  + 3 \Sigma_3^4 \Big) \notag
  \\
  &= (N^2-1) \Big( \frac{N^2-9}{4} \cosh (2\theta) + 2 \theta^2 \cosh \theta 
  - (N^2-5)\cosh \theta + \frac{3N^2-11}{4} \Big) {\bf 1}_N \notag
  \\
  &\quad\,
  - 2 (N^2-3) \big( \sinh (2\theta) - \theta^2 \sinh \theta - 2 \sinh \theta \big) \Sigma_3 \notag
  \\
  &\quad\,
  - \Big( \frac{N^2 \!-\! 11}{2} \cosh (2 \theta) \!-\! \frac{N^2 \!-\! 13}{2} \theta^2 \cosh \theta \!-\! 2 (N^2 \!-\! 5) \cosh \theta
  \!+\! \frac{N^2 \!-\! 1}{2} \theta^2 \!+\! \frac{3 (N^2 \!-\! 3)}{2} \Big) \Sigma_3^2 \notag
  \\
  &\quad\,
  + \big( 2 \sinh (2 \theta) - 3 \theta^2 \sinh \theta - 4 \sinh \theta + \theta^3 \big) \Sigma_3^3 \notag
  \\
  &\quad\,
  + \Big( \frac{1}{4} \cosh (2\theta) - \frac{1}{2} \theta^2 \cosh \theta 
  - \cosh \theta + \frac{1}{8} \theta^4 + \frac{1}{2} \theta^2 
  + \frac{3}{4} \Big) \Sigma_3^4 
  \\
  &\equiv \sum_{n=0}^4 J_{1,n}(\theta) \Sigma_3^n. \notag
\end{alignat}
And by taking the large $N$ limit and using the eq.~(\ref{eq:trN}), 
the integral of $J_1$ part in eq.~(\ref{eq:Veffsph}) is evaluated as
\begin{alignat}{3}
  &\sqrt{\frac{N}{2}} \int_0^\infty \frac{d\theta}{\theta^{9/2}} 
  e^{-\big( \frac{N^2-1}{4} \tilde{r}^2 + \tilde{x}^2 \big)\frac{\theta}{2\tilde{r}\tilde{x}}} \,
  \text{tr}_N \Big[ e^{\theta \frac{\Sigma_3}{2}} J_1(\theta) \Big] \notag
  \\
  &\sim \Big( \frac{N}{2} \Big)^4 \int_0^\infty \frac{d\chi}{\chi^{9/2}} 
  e^{-g\chi} \, \sum_{n=0}^4 N^{n+1} \frac{d^n}{d\chi^n} \Big( \frac{\sinh\chi}{\chi} \Big) J_{1,n}(2\chi/N) \notag
  \\
  &= \frac{\sqrt{\pi}}{2^3} N^5 \big( \sqrt{g+1} - \sqrt{g-1} \big) + \mathcal{O} (N^3). \label{eq:J1part}
\end{alignat}
Here we defined $g = \frac{\tilde{R}_\text{sph}}{2\tilde{x}} + \frac{\tilde{x}}{2\tilde{R}_\text{sph}}$.

Definition of $J_2(\theta)$ and explicit expression are given as follows.
\begin{alignat}{3}
  J_2(\theta) &\equiv \int_0^\theta \!\!\! d\theta_1 \int_0^{\theta_1} \!\!\! d\theta_2
  \int_0^{\theta_2} \!\!\! d\theta_3 \int_0^{\theta_3} \!\!\! d\theta_4 
  \Big( \Sigma_a(\theta_1) \Sigma^a(\theta_2) \Sigma_b(\theta_3) \Sigma^b(\theta_4) \notag
  \\[-0.1cm]
  &\quad\,
  - \Sigma_a(\theta_1) \Sigma_b(\theta_2) \Sigma^b(\theta_3) \Sigma^a(\theta_4) \Big) \notag
  \\
  &= \int_0^\theta \!\!\! d\theta_1 \int_0^{\theta_1} \!\!\! d\theta_2
  \int_0^{\theta_2} \!\!\! d\theta_3 \int_0^{\theta_3} \!\!\! d\theta_4 \Big( 
  4 \sinh (\theta_1 \!+\! \theta_2 \!-\! \theta_3 \!-\! \theta_4) \big( (N^2 \!-\! 3) \Sigma_3 \!-\! \Sigma_3^3 \big) \notag
  \\
  &\quad\,
  - \tfrac{1}{2} \cosh (\theta_1 \!+\! \theta_2 \!-\! \theta_3 \!-\! \theta_4) 
  \big( (N^2 \!-\! 9)(N^2 \!-\! 1) \!-\! 2 (N^2 \!-\! 11) \Sigma_3^2 \!+\! \Sigma_3^4 \big) \notag
  \\
  &\quad\, 
  + \tfrac{1}{2} \cosh (\theta_1 \!-\! \theta_2 \!-\! \theta_3 \!+\! \theta_4) 
  \big( (N^2 \!-\! 1)^2 \!-\! 2 (N^2 \!+\! 1) \Sigma_3^2 \!+\! \Sigma_3^4 \big)
  - 2 \sinh (\theta_1 \!-\! \theta_2) \Sigma_3^3 \notag
  \\
  &\quad\,
  - 2 \sinh (\theta_1 \!-\! \theta_4) \big( 2 (N^2 \!-\! 3) \Sigma_3 \!-\! 3 \Sigma_3^3 \big)
  + 2 \sinh (\theta_2 \!-\! \theta_3) \Sigma_3^3
  - 2 \sinh (\theta_3 \!-\! \theta_4) \Sigma_3^3 \notag
  \\
  &\quad\,
  + \cosh (\theta_1 \!-\! \theta_2) \big( (N^2 \!-\! 1) \Sigma_3^2 \!-\! \Sigma_3^4 \big) 
  - \cosh (\theta_1 \!-\! \theta_4) \big( 4 (N^2 \!-\! 1) + (N^2 \!-\! 13) \Sigma_3^2 \!-\! \Sigma_3^4 \big) \notag
  \\
  &\quad\,
  - \cosh (\theta_2 \!-\! \theta_3) \big( (N^2 \!-\! 1) \Sigma_3^2 \!-\! \Sigma_3^4 \big) 
  + \cosh (\theta_3 \!-\! \theta_4) \big( (N^2 \!-\! 1) \Sigma_3^2 \!-\! \Sigma_3^4 \big) \Big) \notag
  \\
  &= (N^2-1) \Big( - \frac{N^2-9}{8} \cosh (2\theta) - 2 \theta^2 \cosh \theta + \frac{N^2+7}{2} \theta \sinh\theta
  - 8 \cosh \theta - \frac{N^2-1}{4} \theta^2 \notag
  \\
  &\quad\,
  + \frac{N^2+55}{8} \Big) {\bf 1}_N
  + (N^2-3) \big( \sinh (2\theta) - 2 \theta^2 \sinh \theta + 4 \theta \cosh\theta - 8 \sinh \theta + 2 \theta \big) \Sigma_3 \notag
  \\
  &\quad\,
  + \Big( \frac{N^2 \!-\! 11}{4} \cosh (2 \theta) \!-\! \frac{N^2 \!-\! 13}{2} \theta^2 \cosh \theta \!-\! 2 (N^2 \!-\! 13) \cosh \theta
  \!+\! (N^2 \!-\! 15) \theta \sinh\theta \!+\! \theta^2 \notag
  \\
  &\quad\,
  +\! \frac{7N^2 \!-\! 93}{4} \Big) \Sigma_3^2 
  + \Big( - \sinh (2 \theta) + 3 \theta^2 \sinh \theta - 8 \theta \cosh\theta + 12 \sinh \theta 
  + \frac{1}{3} \theta^3 - 2 \theta \Big) \Sigma_3^3 \notag
  \\
  &\quad\,
  + \Big( - \frac{1}{8} \cosh (2\theta) + \frac{1}{2} \theta^2 \cosh \theta - \frac{3}{2} \theta \sinh\theta
  + 2 \cosh \theta + \frac{1}{4} \theta^2 - \frac{15}{8} \Big) \Sigma_3^4
  \\
  &\equiv \sum_{n=0}^4 J_{2,n}(\theta) \Sigma_3^n . \notag
\end{alignat}
And by taking the large $N$ limit and using the eq.~(\ref{eq:trN}), 
the integral of $J_2$ part in eq.~(\ref{eq:Veffsph}) is evaluated as
\begin{alignat}{3}
  &\sqrt{\frac{N}{2}} \int_0^\infty \frac{d\theta}{\theta^{9/2}} 
  e^{-\big( \frac{N^2-1}{4} \tilde{r}^2 + \tilde{x}^2 \big)\frac{\theta}{2\tilde{r}\tilde{x}}} \,
  \text{tr}_N \Big[ e^{\theta \frac{\Sigma_3}{2}} J_2(\theta) \Big] \notag
  \\
  &\sim \Big( \frac{N}{2} \Big)^4 \int_0^\infty \frac{d\chi}{\chi^{9/2}} 
  e^{-g\chi} \, \sum_{n=0}^4 N^{n+1} \frac{d^n}{d\chi^n} \Big( \frac{\sinh\chi}{\chi} \Big) J_{2,n}(2\chi/N) \notag
  \\
  &= - \frac{\sqrt{\pi}}{180} N^3 
  \frac{ (2 g (\sqrt{g-1}-\sqrt{g+1})+ \sqrt{g-1} + \sqrt{g+1} )}{\sqrt{g-1} \sqrt{g+1} } + \mathcal{O} (N).
\end{alignat}
This is subleading compared with the eq.~(\ref{eq:J1part}).


\section{Calculation of $L_1$ and $L_2$} \label{sec:L12}


Definition of $L_1(\ell)$ and explicit expression are given as follows.
\begin{alignat}{3}
  \big( L_1(\ell) \big)_{mn} &\equiv \int_0^\ell \!\!\! d\ell_1 \int_0^{\ell_1} \!\!\! d\ell_2
  \int_0^{\ell_2} \!\!\! d\ell_3 \int_0^{\ell_3} \!\!\! d\ell_4 
  \Big( \Xi_a(\ell_1) \Xi^a(\ell_2) \Xi_b(\ell_3) \Xi^b(\ell_4) \notag
  \\[-0.1cm]
  &\quad\,
  + \Xi_a(\ell_1) \Xi_b(\ell_2) \Xi^b(\ell_3) \Xi^a(\ell_4)
  + \Xi_a(\ell_1) \Xi_b(\ell_2) \Xi^a(\ell_3) \Xi^b(\ell_4) \Big)_{mn} \notag
  \\
  &= \frac{1}{2} \int_0^\ell \!\!\! d\ell_1 \int_0^{\ell_1} \!\!\! d\ell_2 
  \int_0^{\ell_2} \!\!\! d\ell_3 \int_0^{\ell_3} \!\!\! d\ell_4 
  \Big( 
  e^{ (\ell _1-\ell _2+\ell _3-\ell_4 ) \lambda_m }
  + e^{ (- \ell_1 + \ell_2 - \ell_3 + \ell_4 ) \lambda_{m-1} }\notag
  \\[0.1cm]
  &\quad\,
  + e^{ (\ell_1 - \ell_4) \lambda_m + (\ell _2 - \ell _3) \lambda_{m+1} } 
  + e^{ (- \ell_1 + \ell_2) \lambda_{m-1} + (\ell_3 - \ell_4) \lambda_m } \notag
  \\
  &\quad\,
  + e^{ (\ell_1 - \ell_2) \lambda_m - (\ell_3 - \ell_4 ) \lambda_{m-1} }
  + e^{ (- \ell_1 + \ell_4) \lambda_{m-1} - (\ell_2 - \ell_3) \lambda_{m-2} } \Big) \delta_{mn} \notag
  \\
  &= \Big\{ \tfrac{(\lambda_{m-1} - \lambda_m)^2}{4 \lambda_{m-1}^2 \lambda_m^2} \ell^2 
  - \tfrac{\lambda_{m-2} - \lambda_{m-1}}{2 \lambda_{m-2} \lambda_{m-1}^3} \ell e^{- \ell \lambda_{m-1}}
  - \tfrac{\lambda_m - \lambda_{m+1}}{2 \lambda_m^3 \lambda_{m+1}} \ell e^{\ell \lambda_m} \notag
  \\
  &\quad\,
  + \Big( \tfrac{(\lambda_{m-1} - \lambda_m) (\lambda_{m-1}^2 + \lambda_m^2)}{\lambda_{m-1}^3 \lambda_m^3} 
  + \tfrac{1}{2 \lambda_{m-1}^2 (\lambda_{m-2} + \lambda_{m-1})}
  - \tfrac{1}{2 \lambda_m^2 (\lambda_m + \lambda_{m+1})} \Big) \ell \notag
  \\
  &\quad\,
  + \Big( \tfrac{1}{\lambda_{m-1}^3 (\lambda_{m-1} + \lambda_m)} 
  - \tfrac{3 \lambda_{m-2}^2 - 2 \lambda_{m-1} \lambda_{m-2} + \lambda_{m-1}^2}{2 \lambda_{m-2}^2 \lambda_{m-1}^4} \Big) 
  e^{-\ell \lambda_{m-1}} 
  \\
  &\quad\,
  - \Big( \tfrac{3 \lambda_{m-1} + \lambda_m}{2 \lambda_m^4 (\lambda_{m-1} + \lambda_m)}
  + \tfrac{\lambda_m - 2 \lambda_{m+1}}{2 \lambda_m^3 \lambda_{m+1}^2} \Big) e^{\ell \lambda_m}
  + \tfrac{e^{- \ell (\lambda_{m-2} + \lambda_{m-1})}}{2 \lambda_{m-2}^2 (\lambda_{m-2} + \lambda_{m-1})^2} 
  + \tfrac{e^{\ell (\lambda_m + \lambda_{m+1})}}{2 \lambda_{m+1}^2 (\lambda_m + \lambda_{m+1})^2} \notag
  \\
  &\quad\,
  + \Big( \tfrac{\lambda_{m-2} (3 \lambda_{m-2} + 4 \lambda_{m-1})}{2\lambda_{m-1}^4 (\lambda_{m-2} + \lambda_{m-1})^2}
  + \tfrac{3 \lambda_{m-1}^3 - 2 \lambda_m \lambda_{m-1}^2 + 2 \lambda_m^2 \lambda_{m-1} - 2 \lambda_m^3}{2 \lambda_{m-1}^3 \lambda_m^4}
  - \tfrac{3 \lambda_m+2 \lambda_{m+1}}{2 \lambda_m^3 (\lambda _m + \lambda_{m+1})^2} \Big) \Big\} \delta_{mn}. \notag
\end{alignat}
And by using the limit (\ref{eq:tr2}), the integral of the $L_1$ part in the eq.~(\ref{eq:Veffcyl}) is evaluated as
\begin{alignat}{3}
  &\int_0^\infty \frac{d\ell}{\ell^{9/2}} e^{- \ell \tilde{\rho}^2 } \,
  \text{tr}_N \Big[ e^{- \ell \tilde{l}^2 \big( \Xi_3 - \frac{\tilde{x}}{\tilde{l}} {\bf 1} \big)^2 } \, L_1(\ell) \Big] \notag
  \\
  &= \sqrt{\pi} \int_{-\infty}^\infty d\zeta \frac{1}{210 \, \tilde{l}} 
  \bigg\{ \frac{140 \big( \zeta^2 + \frac{\tilde{\rho}^2}{\tilde{l}^2} \big)^{3/2}}{(1 - 4 \zeta^2)^2}
  - \frac{14 (32 \zeta^4 - 44 \zeta^2 - 15) 
  \big( \zeta^2 + \frac{\tilde{\rho}^2}{\tilde{l}^2} \big)^{5/2}}{(\zeta^2-1) (4 \zeta^2 - 1)^3} \notag
  \\
  &\qquad\qquad\qquad\quad
  + \frac{3 (64 \zeta^8 - 144 \zeta^6 - 244 \zeta^4 + 241 \zeta^2 + 29) 
  \big( \zeta^2 + \frac{\tilde{\rho}^2}{\tilde{l}^2} \big)^{7/2}}{(1-4\zeta^2)^4 (\zeta^2-1)^2} \label{eq:L1hat}
  \\
  &\qquad\qquad\qquad\quad
  - \frac{8 (8 \zeta^3 + 20 \zeta^2 + 14 \zeta -9)
  \big( (\zeta +1)^2 + \frac{\tilde{\rho}^2}{\tilde{l}^2} \big)^{7/2}}{\zeta (2 \zeta +1)^4 (2 \zeta +3)^2} \notag
  \\
  &\qquad\qquad\qquad\quad
  +\frac{\big((\zeta +2)^2 + \frac{\tilde{\rho}^2}{\tilde{l}^2} \big)^{7/2}}{(\zeta +1)^2 (2\zeta +3)^2} 
  + \frac{112 \big((\zeta +1)^2 + \frac{\tilde{\rho}^2}{\tilde{l}^2} \big)^{5/2}}{(2 \zeta +1)^3 (2 \zeta +3)} 
  + \big(\zeta \to - \zeta \big) \bigg\} \notag
  \\
  &\equiv \sqrt{\pi} \int_{-\infty}^\infty d\zeta \; \hat{L}_1 \big( \zeta,\tfrac{\tilde{\rho}}{\tilde{l}} \big). \notag
\end{alignat}

Definition of $L_2(\ell)$ and explicit expression are given as follows.
\begin{alignat}{3}
  \big(L_2(\ell)\big)_{mn} &\equiv \int_0^\ell \!\!\! d\ell_1 \int_0^{\ell_1} \!\!\! d\ell_2
  \int_0^{\ell_2} \!\!\! d\ell_3 \int_0^{\ell_3} \!\!\! d\ell_4 
  \Big( \Xi_a(\ell_1) \Xi^a(\ell_2) \Xi_b(\ell_3) \Xi^b(\ell_4) \notag
  \\[-0.1cm]
  &\quad\,
  - \Xi_a(\ell_1) \Xi_b(\ell_2) \Xi^b(\ell_3) \Xi^a(\ell_4) \Big)_{mn} \notag
  \\
  &= \Big\{ - \tfrac{1}{4 \lambda_{m-1} \lambda_m} \ell^2
  - \tfrac{\lambda_{m-1} - \lambda_m}{2\lambda_{m-1}^2 \lambda_m^2} \ell 
  + \tfrac{1}{2 \lambda_{m-1}^3 (\lambda_{m-1} + \lambda_m)} e^{-\ell  \lambda_{m-1}} \notag
  \\
  &\quad\,
  + \tfrac{1}{2 \lambda_m^3 (\lambda_{m-1} + \lambda_m)} e^{\ell \lambda_m}
  - \tfrac{\lambda_{m-1}^2 - \lambda_m \lambda_{m-1} + \lambda_m^2}{2 \lambda_{m-1}^3 \lambda_m^3} \Big\} \delta_{mn}.
\end{alignat}
And by using the limit (\ref{eq:tr2}), the integral of the $L_2$ part in the eq.~(\ref{eq:Veffcyl}) is evaluated as
\begin{alignat}{3}
  &\int_0^\infty \frac{d\ell}{\ell^{9/2}} e^{- \ell \tilde{\rho}^2 } \,
  \text{tr}_N \Big[ e^{- \ell \tilde{l}^2 \big( \Xi_3 - \frac{\tilde{x}}{\tilde{l}} {\bf 1} \big)^2 } \, L_2(\ell) \Big] \notag
  \\
  &= \sqrt{\pi} \int_{-\infty}^\infty d\zeta \frac{1}{420 \, \tilde{l}}
  \bigg\{ - \frac{32 (4 \zeta^2 + 3) \big(\zeta^2 + \frac{\tilde{\rho}^2}{\tilde{l}^2} \big)^{7/2}}{(4 \zeta^2 - 1)^3}
  + \frac{224 \big(\zeta^2 + \frac{\tilde{\rho}^2}{\tilde{l}^2} \big)^{5/2}}{(4 \zeta^2 - 1)^2} 
  \\
  &\qquad\qquad\qquad\quad\;\,
  - \frac{140 \big( \zeta^2 + \frac{\tilde{\rho}^2}{\tilde{l}^2} \big)^{3/2}}{4 \zeta^2 - 1}
  + \frac{8 \big((\zeta -1)^2 + \frac{\tilde{\rho}^2}{\tilde{l}^2} \big)^{7/2}}{\zeta (2 \zeta - 1)^3}
  + \frac{8 \big((\zeta +1)^2 + \frac{\tilde{\rho}^2}{\tilde{l}^2} \big)^{7/2}}{\zeta (2 \zeta + 1)^3} \bigg\} \notag
  \\
  &\equiv \sqrt{\pi} \int_{-\infty}^\infty d\zeta \; \hat{L}_2 \big( \zeta,\tfrac{\tilde{\rho}}{\tilde{l}} \big). \notag
\end{alignat}


\end{document}